\documentclass[aps,prb,twocolumn,groupedaddress, showpacs]{revtex4}

\usepackage{graphicx}

\usepackage{hyperref}

\begin{document}

\title{Effects of carrier mobility and morphology in organic semiconductor spin valves}

\author{Yaohua Liu}
\affiliation{Department of Physics and Astronomy, The Johns Hopkins University, Baltimore, MD, 21218}
\author{Taegweon Lee}
\affiliation{Department of Materials Science and Engineering, The Johns Hopkins University, Baltimore, MD, 21218}
\author{Howard E. Katz}
\affiliation{Department of Materials Science and Engineering, The Johns Hopkins University, Baltimore, MD, 21218}
\author{Daniel H. Reich}
\email[Corresponding author, ]{reich@jhu.edu}
\affiliation{Department of Physics and Astronomy, The Johns Hopkins University, Baltimore, MD, 21218}

\date{\today}

\begin{abstract}

We studied spin transport in four organic semiconductors (OSCs) with different electronic properties, with Fe and Co as the top and bottom ferromagnetic (FM) contacts, respectively. Magnetoresistance (MR) effects were observed up to room temperature in junctions based on an electron-carrying OSC, tris(8-hyroxyquinoline) aluminum (Alq$_3$) and a hole-carrying OSC, copper phthalocyanine (CuPc). The MR shows similar temperature dependence for these two OSCs, which suggests that the FM leads rather than the OSCs play a dominant role on the spin-transport degradation with increasing temperature. We also investigated junctions based on two high lateral mobility electron-carrying OSCs, 3,4,9,10-perylenetetracarboxylic dianhydride (PTCDA) and N, N'-bis(4-trifluoromethylbenzyl)-1,4,5,8-naphthalenetetracarboxylic diimide (CF$_3$-NTCDI). However, these junctions showed much weaker spin transport effects. Morphological studies suggest that these high mobility OSC films have much rougher surfaces than either Alq$_3$ or CuPc, therefore the degradation of spin transport may originate from enhanced scattering due to the rougher FM/OSC interfaces. Our study shows that FM/OSC interfaces play an important role for spin transport in organic devices and need further exploration.
\end{abstract}

\maketitle

Spin-dependent electronic transport in organic semiconductors (OSCs) has attracted significant interest recently. The weak spin-orbit coupling in OSCs may lead to long spin relaxation times and diffusion lengths, which is desirable for magnetoelectronic devices. Large magnetoresistance (MR) effects were originally reported in La$_{0.67}$Sr$_{0.33}$MnO$_{3}$ (LSMO)/OSC/Co trilayers, where the OSC was tris(8-hydroxyquinoline) aluminum (Alq$_3$).~\cite{XiongNature2004} Such effects have since been observed in LSMO/OSC/Co systems for a range of OSCs,~\cite{WangPRB2007, OsterbackaAPL2006} and in structures in which both FM layers are transition metals.~\cite{WangSynthMetal2005, SantosPRL2007, ShimPRL2008} However, the observed spin diffusion lengths $\lambda _s$ in the OSCs in such systems are only on the order of 50 nm or less at T $\sim$ 10 K.~\cite{XiongNature2004, ShimPRL2008} Therefore it is of interest to identify new OSCs with potentially enhanced spin transport. One potential reason for the observed short $\lambda _s$ is the low mobility associated with the amorphous OSCs used in previous transport studies. ($\lambda _s = \mu E \tau _s$, where $\tau_s$ is the spin relaxation time, $\mu$ is the carriers' mobility and $E$ is the electric field strength.) However, the mobility of OSCs spans many orders of magnitude, depending on both chemical structure and morphology, and longer spin diffusion lengths may be expected in OSCs with higher mobility. In addition, although Alq$_3$ is an electron transporter, most air-stable high-mobility OSCs are hole transporters and thus it is worth studying spin transport in hole-carrying OSCs, as implementation of spin transport for both hole and electron carriers will allow more freedom to tailor device performance.

To explore these issues, we measured the magnetotransport properties of Fe/OSC/Co multilayer junctions using four different OSC interlayers with different electronic properties. These included three electron-carrying OSCs: Alq$_3$, 3,4,9,10-perylenetetracarboxylic dianhydride (PTCDA) and N, N'-bis(4-trifluoromethylbenzyl)-1,4,5,8-naphthalenetetracarboxylic diimide (CF$_3$-NTCDI), and one hole-carrying OSC, copper phthalocyanine (CuPc). The mobilities in these OSCs are of the order of 10$^{-6}$, 10$^{-4}$, 10$^{-1}$ and 10$^{-3}$ cm$^2$/V$\cdot$S respectively at room temperature.~\cite{ParkAPL2007, JoyceJACS1996, Katz2001, BaoAPL1996} Co and Fe were used as the top and bottom ferromagnetic (FM) layers respectively. MR effects were observed in junctions based on all four OSCs. For junctions based on Alq$_3$ and CuPc, large MRs were observed at T = 80 K, and these effects persisted to room temperature with similar temperature dependence. This similarity suggests that the origin of the temperature dependence is more likely related to the FM leads rather than the OSCs.~\cite{WangPRB2007} In contrast, we found that the junctions based on PTCDA and CF$_3$-NTCDI showed much weaker MR. While the large lateral mobilities in these materials may not be indicative of large mobilities in vertical devices, AFM studies show that both of these OSCs have much rougher surfaces than Alq$_3$ and CuPc. The degradation of spin transport may therefore be due to increased scattering for charge carriers of both spin orientations at the rougher FM/OSC interfaces.~\cite{SchadPRB1998}

Fe/OSC/Co junctions were prepared in thermal evaporation chambers with a base pressure of 2 $\mu$Torr. The samples were deposited on Si wafers with 300 nm SiO$_2$ top layers and had the structure 25 nm Fe/$\sim$ 100 nm OSC/5 nm Co /40 nm Al, using either cross junction~\cite{XiongNature2004} or overlap geometries.~\cite{overlap} The Al cap layer served to prevent oxidation of the Co layer. In order to stabilize the OSC films during evaporation and to reduce the potential penetration of Co atoms into the OSC, chilled water was used to keep the substrate holder at 20 $^{\circ}$C during the Co and Al deposition. The transport measurements for the Alq$_3$ and CuPc junctions were carried out in a continuous flow cryostat in the temperature range 40 K $\leq$ T $\leq$ 290 K. An electromagnet was used to provide an in-plane magnetic field parallel to the top Co electrode to control the relative magnetization direction of the top and bottom FM layers. The PTCDA and CF$_3$-NTCDI devices were measured at 4 K in a cryostat equipped with a superconducting magnet. The surface morphology of OSC films on top of Fe layers was studied via atomic force microscopy (AFM), using tapping mode.

\begin{figure}
	\centering
		\includegraphics{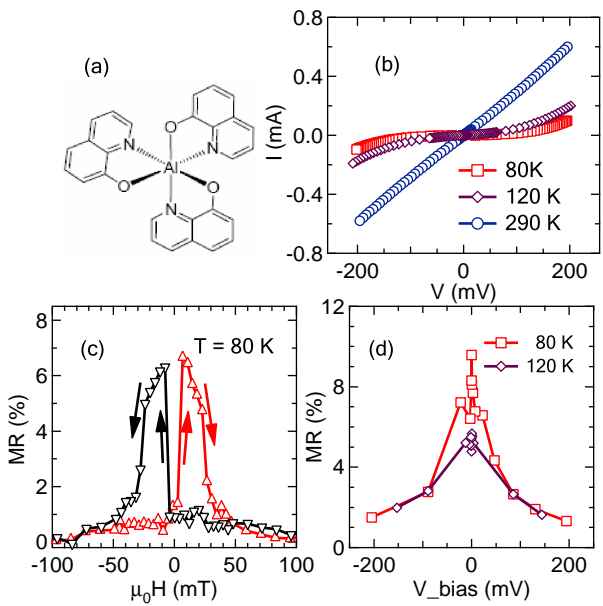}
	\caption{\label{Alq}(a) Chemical structure of Alq$_3$. (b) I-V curves taken at T = 80 K, 120 K and 290 K for a Fe/Alq$_3$/Co junction. (c) Magnetotransport data at T = 80 K with I$_{bias}$ = 1 $\mu$A. (d) Bias dependence of MR taken at T = 80 K and 120 K. The maximum observed MR is 9.5\% at T = 80 K and 5.7\% at T = 120 K.}
\end{figure}

Figure~\ref{Alq}a shows the chemical structure of Alq$_3$. Figure~\ref{Alq}b shows current-voltage (I-V) curves taken at T = 80 K, 120 K and 290 K for an Alq$_3$-based junction. These are distinctly nonlinear at low temperatures, but become more linear as the temperature increases. The I-V curves show very weak asymmetry for positive bias and negative bias, which is due to the similar barriers that are created at Fe/OSC and Co/OSC interfaces when fabricated at base pressures of $\sim$ 1 $\mu$Torr.~\cite{CambellAPL2007} Both the nonlinear I-V curves and the temperature dependence of the junction resistance exclude the possibility of a metallic short between the two FM layers. A magnetotransport curve taken at T = 80 K is shown in Fig.~\ref{Alq}c, where 6.6\% MR is observed. The bias dependence of the MR at T = 80 K and 120 K are shown in Fig.~\ref{Alq}d. The maximum observed MR was 9.5\% at T = 80 K with I$_{bias}$ = 10 nA. The MR decreased with increasing bias at both temperatures. This tendency has been generally observed in previous studies, and has been attributed to the bias-dependent polarization of the FM leads or the excitation of magnons or phonons at higher voltages.~\cite{XiongNature2004, WangPRB2007, SantosPRL2007} In our case, the bias dependence became weaker at higher temperature, which suggests that bias-dependent polarization of the FM leads is not the only mechanism here.

\begin{figure}
	\centering
		\includegraphics{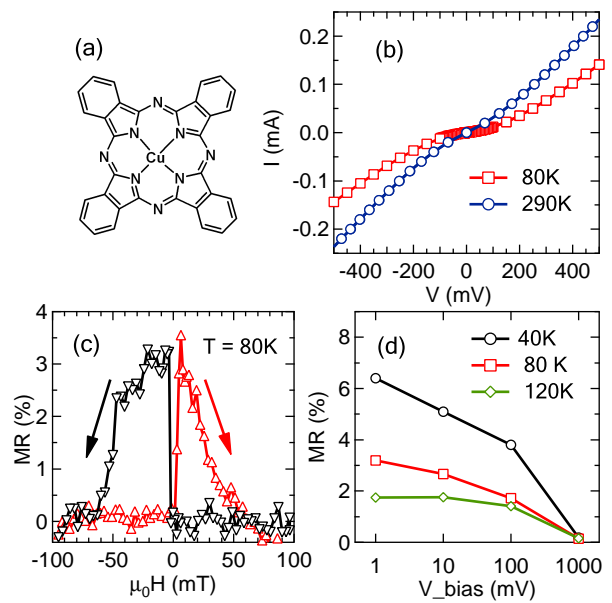}
	\caption{\label{CuPc} (a) Chemical structure of CuPc. (b) I-V curves at T = 80 K and 290 K for a CuPc-based device. (c) Magnetotransport data taken at T = 80 K with V$_{bias}$ = 1 mV. (d) Bias dependence of MR at T = 40 K, 80 K and 120 K. Maximum MRs are 6.4\%, 3.2\% and 1.8\% at these three temperatures, respectively.}
\end{figure}

We also studied the spin transport in the hole-carrying OSC CuPc. The chemical structure of CuPc is shown in Fig.~\ref{CuPc}a. Figure~\ref{CuPc}b shows I-V curves taken at T = 80 K and 290 K, which show a weaker temperature dependence than those of the Alq$_3$-based device, but are also nearly symmetric for positive and negative bias. Figures~\ref{CuPc}c and \ref{CuPc}d show the magnetotransport results. Strong bias-dependence was also observed in these CuPc-based devices, with decreased amplitude at higher temperatures.

\begin{figure}
	\centering
		\includegraphics{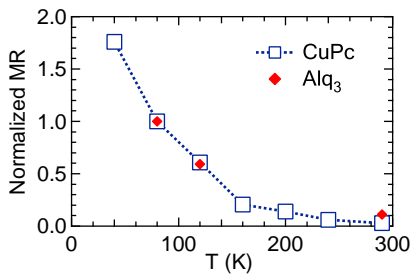}
	\caption{\label{Temp} Temperature dependence of maximum observed MR for two junctions based on Alq$_3$ and CuPc, respectively. The MRs for each sample are normalized to their values at T = 80 K.}
\end{figure}

For junctions based on both Alq$_3$ and CuPc, the MR persists to room temperature, although with decreased amplitude. Insight into the origin of the temperature dependence may be obtained from a comparison of the results for Alq$_3$- and CuPc-based junctions, as shown in Fig.~\ref{Temp}. When the maximum MRs are normalized to their values at 80 K, the two sets of data nearly overlap. This similarity suggests that the temperature dependence is not governed by the OSCs.~\cite{WangPRB2007} Wang \emph{et al.} observed the same similarity in different OSCs and attributed the temperature dependence to decreasing surface spin polarization of LSMO with increasing temperature.~\cite{WangPRB2007} Although we used high Curie temperature FM leads, the MR in our samples still shows strong temperature dependence below room temperature, which suggests that the spin polarization may have a stronger temperature dependence at the FM surface than that of the bulk FM materials.~\cite{WangPRB2007, WalkerJAP1984, Taborelli1988} It has been observed that the surface magnetization of Fe films has a much stronger temperature dependence than the bulk magnetization and that this dependence is also related to the material adjacent to the Fe film.~\cite{WalkerJAP1984, Taborelli1988} Therefore further study is warranted on the surface magnetization of these transition metallic films when they are adjacent to OSCs to elucidate the temperature dependence of the MR.

\begin{figure}
	\centering
		\includegraphics{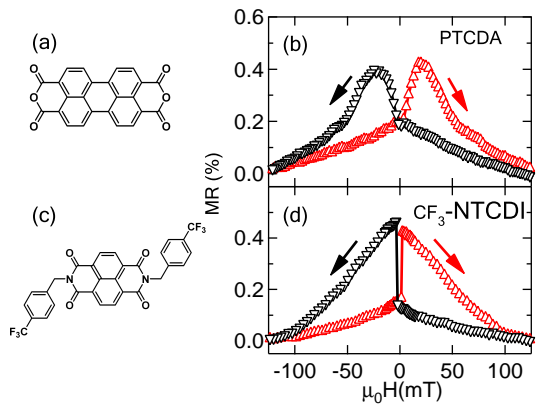}
	\caption{\label{PTCDA_NTCDI}(a), (c) Chemical structures of PTCDA and CF$_3$-NTCDI. (b), (d) Magnetotransport data taken at T = 4.2 K, for junctions based on PTCDA and CF$_3$-NTCDI respectively. }
\end{figure}

\begin{figure}
	\centering
		\includegraphics{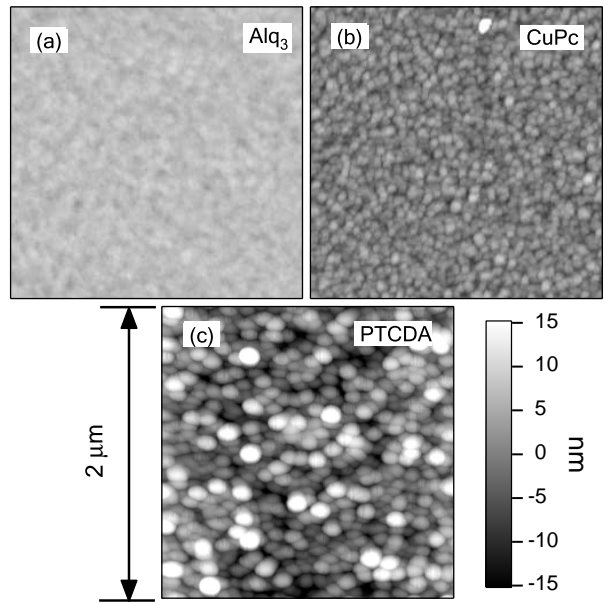}
	\caption{\label{AFM}(a)-(c) AFM images of Alq$_3$, CuPc  and PTCDA films on top of 25 nm Fe layers respectively. The images were scanned over 2 $\times$ 2 $\mu$m$^2$ areas via tapping mode. The PTCDA film has a much rougher surface than the Alq$_3$ and CuPc films.}
\end{figure}

Figure~\ref{PTCDA_NTCDI} shows magnetotransport data for two junctions based on the high mobility OSCs PTCDA and CF$_3$-NTCDI. The transport data were acquired at T = 4.2 K. Less than 1\% MR was observed for each OSC, which is much smaller than that in Alq$_3$-based junctions. This observation is contrary to the expectation that enhanced spin transport happens in higher mobility OSCs due to longer spin diffusion length. It is worth noting, however, that the mobility values for these materials were determined from measurement in lateral geometries~\cite{BaoAPL1996, Katz2001} and may not reflect the mobilities perpendicular to the film plane, which are relevant for our measurements. This may be true in particular for CF$_3$-NTCDI, which has insulating side chains. However, in order to achieve effective spin transport, spin polarized current must first be injected into the OSC, and the FM/OSC interfaces are critical for this process. We compared the surface structure of these OSCs via AFM. Figure~\ref{AFM} shows the surface morphology for Alq$_3$, CuPc and PTCDA films on top of 25 nm Fe layers. It is clear that PTCDA forms a film with a significantly rougher surface than either Alq$_3$ or CuPc. Therefore the degradation of spin transport is associated with rougher OSC films. This is likely because enhanced scattering at the rough interface can cause strong spin depolarization.~\cite{SchadPRB1998} Our observation that larger MR associated with smoother OSC surfaces is contrary to the results reported by Xu \emph{et al.}~\cite{XuAPL2007} While they attributed the observed MR to tunneling through hotspots caused by rough OSC films, our results suggest that a smooth interface is very important for effective spin injection.

Charge injection barriers also play a very important role in achieving effective spin injection from FM metals into OSCs,~\cite{RudenSmithJAP2004, DediuPRB2008} and barriers larger than 0.6 eV are typically required.~\cite{RudenSmithJAP2004, CambellAPL2007} Previous photoemission spectroscopies have shown the breakdown of the vacuum level alignment rule at interfaces between OSCs and metals.~\cite{HillAPL1998, KahnPolymer2003, TibaJAP2006,PopinciucJAP2006, ZhanPRB2007, ZhanPRB2008} Experimental evidence shows that Alq$_3$ and CuPc form larger charge injection barriers (about 1 eV) when adjacent to metals with work functions close to that of Co and Fe.~\cite{HillAPL1998, KahnPolymer2003, ZhanPRB2008} However charge injection barriers from metals to PTCDA are almost independent of the metals, and are only about 0.3 eV.~\cite{HillAPL1998} These barrier values seems consistent with our observation of larger MR in Alq$_3$ and CuPc-based junctions. However this picture of large charge injection barriers does not explain the observation of MR at very low bias, which suggests other mechanisms may govern the spin injection in this regime.~\cite{ZhanPRB2008}

Our work demonstrates that it is possible to inject spin polarized current into both electron- and hole-carrying OSCs at room temperature from high Curie temperature transition metals. The similar temperature dependence of the MR for Alq$_3$- and CuPc-based junctions suggests that the degradation of spin transport with increasing temperature may originate from decreased surface spin polarization of the FM leads at the OSC/FM interfaces. Although the variation in charge barriers may also play a role, the high mobility OSCs we have studied, PTCDA and CF$_3$-NTCDI form films with very rough surfaces, which gives a possible explanation for the observation of much smaller MR. Our work suggests that FM/OSC interfaces play an important role for spin transport. A sharper FM/OSC interface with weak temperature-dependent surface polarization appears necessary to improve spin transport in OSCs at room temperature. Further studies are still needed to elucidate the role of interfaces in these systems.

This work was supported by NSF Grant No. DMR-0520491.


\begin{thebibliography}{24}
\expandafter\ifx\csname natexlab\endcsname\relax\def\natexlab#1{#1}\fi
\expandafter\ifx\csname bibnamefont\endcsname\relax
  \def\bibnamefont#1{#1}\fi
\expandafter\ifx\csname bibfnamefont\endcsname\relax
  \def\bibfnamefont#1{#1}\fi
\expandafter\ifx\csname citenamefont\endcsname\relax
  \def\citenamefont#1{#1}\fi
\expandafter\ifx\csname url\endcsname\relax
  \def\url#1{\texttt{#1}}\fi
\expandafter\ifx\csname urlprefix\endcsname\relax\def\urlprefix{URL }\fi
\providecommand{\bibinfo}[2]{#2}
\providecommand{\eprint}[2][]{\url{#2}}

\bibitem[{\citenamefont{Xiong et~al.}(2004)\citenamefont{Xiong, Wu, Vardeny,
  and Shi}}]{XiongNature2004}
\bibinfo{author}{\bibfnamefont{Z.~H.} \bibnamefont{Xiong}},
  \bibinfo{author}{\bibfnamefont{D.}~\bibnamefont{Wu}},
  \bibinfo{author}{\bibfnamefont{Z.~V.} \bibnamefont{Vardeny}},
  \bibnamefont{and} \bibinfo{author}{\bibfnamefont{J.}~\bibnamefont{Shi}},
  \bibinfo{journal}{Nature} \textbf{\bibinfo{volume}{427}},
  \bibinfo{pages}{821} (\bibinfo{year}{2004}).

\bibitem[{\citenamefont{Wang et~al.}(2007)\citenamefont{Wang, Yang, Vardeny,
  and Li}}]{WangPRB2007}
\bibinfo{author}{\bibfnamefont{F.~J.} \bibnamefont{Wang}},
  \bibinfo{author}{\bibfnamefont{C.~G.} \bibnamefont{Yang}},
  \bibinfo{author}{\bibfnamefont{Z.~V.} \bibnamefont{Vardeny}},
  \bibnamefont{and} \bibinfo{author}{\bibfnamefont{X.~G.} \bibnamefont{Li}},
  \bibinfo{journal}{Phys. Rev. B} \textbf{\bibinfo{volume}{75}},
  \bibinfo{pages}{245324} (\bibinfo{year}{2007}).

\bibitem[{\citenamefont{Majumdara et~al.}(2006)\citenamefont{Majumdara, Laiho,
  Laukkanen, V{\"{a}}ryne, Majumdar, and {\"{O}}sterbacka}}]{OsterbackaAPL2006}
\bibinfo{author}{\bibfnamefont{S.}~\bibnamefont{Majumdara}},
  \bibinfo{author}{\bibfnamefont{R.}~\bibnamefont{Laiho}},
  \bibinfo{author}{\bibfnamefont{P.}~\bibnamefont{Laukkanen}},
  \bibinfo{author}{\bibfnamefont{I.~J.} \bibnamefont{V{\"{a}}ryne}},
  \bibinfo{author}{\bibfnamefont{H.~S.} \bibnamefont{Majumdar}},
  \bibnamefont{and}
  \bibinfo{author}{\bibfnamefont{R.}~\bibnamefont{{\"{O}}sterbacka}},
  \bibinfo{journal}{Appl. Phys. Lett.} \textbf{\bibinfo{volume}{89}},
  \bibinfo{pages}{122114} (\bibinfo{year}{2006}).

\bibitem[{\citenamefont{Wang et~al.}(2005)\citenamefont{Wang, Xiong, Wu, Shi,
  and Vardeny}}]{WangSynthMetal2005}
\bibinfo{author}{\bibfnamefont{F.}~\bibnamefont{Wang}},
  \bibinfo{author}{\bibfnamefont{Z.}~\bibnamefont{Xiong}},
  \bibinfo{author}{\bibfnamefont{D.}~\bibnamefont{Wu}},
  \bibinfo{author}{\bibfnamefont{J.}~\bibnamefont{Shi}}, \bibnamefont{and}
  \bibinfo{author}{\bibfnamefont{Z.}~\bibnamefont{Vardeny}},
  \bibinfo{journal}{Synth. Met.} \textbf{\bibinfo{volume}{155}},
  \bibinfo{pages}{172} (\bibinfo{year}{2005}).

\bibitem[{\citenamefont{Santos et~al.}(2007)\citenamefont{Santos, Lee, Migdal,
  Lekshmi, Satpati, and Moodera}}]{SantosPRL2007}
\bibinfo{author}{\bibfnamefont{T.~S.} \bibnamefont{Santos}},
  \bibinfo{author}{\bibfnamefont{J.~S.} \bibnamefont{Lee}},
  \bibinfo{author}{\bibfnamefont{P.}~\bibnamefont{Migdal}},
  \bibinfo{author}{\bibfnamefont{I.~C.} \bibnamefont{Lekshmi}},
  \bibinfo{author}{\bibfnamefont{B.}~\bibnamefont{Satpati}}, \bibnamefont{and}
  \bibinfo{author}{\bibfnamefont{J.~S.} \bibnamefont{Moodera}},
  \bibinfo{journal}{Phys. Rev. Lett.} \textbf{\bibinfo{volume}{98}},
  \bibinfo{pages}{016601} (\bibinfo{year}{2007}).

\bibitem[{\citenamefont{Shim et~al.}(2008)\citenamefont{Shim, Raman, Park,
  Santos, Miao, Satpati, and Moodera}}]{ShimPRL2008}
\bibinfo{author}{\bibfnamefont{J.~H.} \bibnamefont{Shim}},
  \bibinfo{author}{\bibfnamefont{K.}~\bibnamefont{Raman}},
  \bibinfo{author}{\bibfnamefont{Y.~J.} \bibnamefont{Park}},
  \bibinfo{author}{\bibfnamefont{T.~S.} \bibnamefont{Santos}},
  \bibinfo{author}{\bibfnamefont{G.~X.} \bibnamefont{Miao}},
  \bibinfo{author}{\bibfnamefont{B.}~\bibnamefont{Satpati}}, \bibnamefont{and}
  \bibinfo{author}{\bibfnamefont{J.~S.} \bibnamefont{Moodera}},
  \bibinfo{journal}{Phys. Rev. Lett.} \textbf{\bibinfo{volume}{100}},
  \bibinfo{pages}{226603} (\bibinfo{year}{2008}).

\bibitem[{\citenamefont{Park et~al.}(2007)\citenamefont{Park, Shin, Yu, and
  Chaea}}]{ParkAPL2007}
\bibinfo{author}{\bibfnamefont{H.}~\bibnamefont{Park}},
  \bibinfo{author}{\bibfnamefont{D.-S.} \bibnamefont{Shin}},
  \bibinfo{author}{\bibfnamefont{H.-S.} \bibnamefont{Yu}}, \bibnamefont{and}
  \bibinfo{author}{\bibfnamefont{H.-B.} \bibnamefont{Chaea}},
  \bibinfo{journal}{Appl. Phys. Lett.} \textbf{\bibinfo{volume}{90}},
  \bibinfo{pages}{202103} (\bibinfo{year}{2007}), \bibinfo{note}{and references
  therein}.

\bibitem[{\citenamefont{Laquindanum et~al.}(1996)\citenamefont{Laquindanum,
  Katz, Dodabalapur, and Lovinger}}]{JoyceJACS1996}
\bibinfo{author}{\bibfnamefont{J.~G.} \bibnamefont{Laquindanum}},
  \bibinfo{author}{\bibfnamefont{H.~E.} \bibnamefont{Katz}},
  \bibinfo{author}{\bibfnamefont{A.}~\bibnamefont{Dodabalapur}},
  \bibnamefont{and} \bibinfo{author}{\bibfnamefont{A.~J.}
  \bibnamefont{Lovinger}}, \bibinfo{journal}{J. Am. Chem. Soc.}
  \textbf{\bibinfo{volume}{118}}, \bibinfo{pages}{11331}
  (\bibinfo{year}{1996}).

\bibitem[{\citenamefont{Katz et~al.}(2001)\citenamefont{Katz, Siegrist,
  Sch{\"o}n, Kloc, Batlogg, Lovinger, and Johnson}}]{Katz2001}
\bibinfo{author}{\bibfnamefont{H.~E.} \bibnamefont{Katz}},
  \bibinfo{author}{\bibfnamefont{T.}~\bibnamefont{Siegrist}},
  \bibinfo{author}{\bibfnamefont{J.~H.} \bibnamefont{Sch{\"o}n}},
  \bibinfo{author}{\bibfnamefont{C.}~\bibnamefont{Kloc}},
  \bibinfo{author}{\bibfnamefont{B.}~\bibnamefont{Batlogg}},
  \bibinfo{author}{\bibfnamefont{A.~J.} \bibnamefont{Lovinger}},
  \bibnamefont{and} \bibinfo{author}{\bibfnamefont{J.}~\bibnamefont{Johnson}},
  \bibinfo{journal}{Chemphyschem} \textbf{\bibinfo{volume}{3}},
  \bibinfo{pages}{167} (\bibinfo{year}{2001}).

\bibitem[{\citenamefont{Bao et~al.}(1990)\citenamefont{Bao, Lovinger, and
  Dodabalapur}}]{BaoAPL1996}
\bibinfo{author}{\bibfnamefont{Z.}~\bibnamefont{Bao}},
  \bibinfo{author}{\bibfnamefont{A.~J.} \bibnamefont{Lovinger}},
  \bibnamefont{and}
  \bibinfo{author}{\bibfnamefont{A.}~\bibnamefont{Dodabalapur}},
  \bibinfo{journal}{Appl. Phys. Lett.} \textbf{\bibinfo{volume}{69}},
  \bibinfo{pages}{11} (\bibinfo{year}{1990}).

\bibitem[{\citenamefont{Schad et~al.}(1998)\citenamefont{Schad, Beli{\"{e}}n,
  Verbanck, Pott, Fischer, Lefebvre, Bessiere, Moshchalkov, and
  Bruynseraede}}]{SchadPRB1998}
\bibinfo{author}{\bibfnamefont{R.}~\bibnamefont{Schad}},
  \bibinfo{author}{\bibfnamefont{P.}~\bibnamefont{Beli{\"{e}}n}},
  \bibinfo{author}{\bibfnamefont{G.}~\bibnamefont{Verbanck}},
  \bibinfo{author}{\bibfnamefont{C.~D.} \bibnamefont{Pott}},
  \bibinfo{author}{\bibfnamefont{H.}~\bibnamefont{Fischer}},
  \bibinfo{author}{\bibfnamefont{S.}~\bibnamefont{Lefebvre}},
  \bibinfo{author}{\bibfnamefont{M.}~\bibnamefont{Bessiere}},
  \bibinfo{author}{\bibfnamefont{V.~V.} \bibnamefont{Moshchalkov}},
  \bibnamefont{and}
  \bibinfo{author}{\bibfnamefont{Y.}~\bibnamefont{Bruynseraede}},
  \bibinfo{journal}{Phys. Rev. B} \textbf{\bibinfo{volume}{57}},
  \bibinfo{pages}{13692} (\bibinfo{year}{1998}).

\bibitem[{\citenamefont{Scheuermann et~al.}(1983)\citenamefont{Scheuermann,
  Lhota, Kuo, and Chen}}]{overlap}
\bibinfo{author}{\bibfnamefont{M.}~\bibnamefont{Scheuermann}},
  \bibinfo{author}{\bibfnamefont{J.~R.} \bibnamefont{Lhota}},
  \bibinfo{author}{\bibfnamefont{P.~K.} \bibnamefont{Kuo}}, \bibnamefont{and}
  \bibinfo{author}{\bibfnamefont{J.~T.} \bibnamefont{Chen}},
  \bibinfo{journal}{Phys. Rev. Lett.} \textbf{\bibinfo{volume}{50}},
  \bibinfo{pages}{74} (\bibinfo{year}{1983}).

\bibitem[{\citenamefont{Campbell and Crone}(2007)}]{CambellAPL2007}
\bibinfo{author}{\bibfnamefont{I.~H.} \bibnamefont{Campbell}} \bibnamefont{and}
  \bibinfo{author}{\bibfnamefont{B.~K.} \bibnamefont{Crone}},
  \bibinfo{journal}{Appl. Phys. Lett.} \textbf{\bibinfo{volume}{90}},
  \bibinfo{pages}{242107} (\bibinfo{year}{2007}).

\bibitem[{\citenamefont{Walker et~al.}(1984)\citenamefont{Walker, Droste,
  Stern, and Tyson}}]{WalkerJAP1984}
\bibinfo{author}{\bibfnamefont{J.~C.} \bibnamefont{Walker}},
  \bibinfo{author}{\bibfnamefont{R.}~\bibnamefont{Droste}},
  \bibinfo{author}{\bibfnamefont{G.}~\bibnamefont{Stern}}, \bibnamefont{and}
  \bibinfo{author}{\bibfnamefont{J.}~\bibnamefont{Tyson}}, \bibinfo{journal}{J.
  Appl. Phys.} \textbf{\bibinfo{volume}{55}}, \bibinfo{pages}{2500}
  (\bibinfo{year}{1984}).

\bibitem[{\citenamefont{Taborelli et~al.}(1988)\citenamefont{Taborelli, Paul,
  Z{\"u}ger, and Landolt}}]{Taborelli1988}
\bibinfo{author}{\bibfnamefont{M.}~\bibnamefont{Taborelli}},
  \bibinfo{author}{\bibfnamefont{O.}~\bibnamefont{Paul}},
  \bibinfo{author}{\bibfnamefont{O.}~\bibnamefont{Z{\"u}ger}},
  \bibnamefont{and} \bibinfo{author}{\bibfnamefont{M.}~\bibnamefont{Landolt}},
  \bibinfo{journal}{J. Phys. Colloques} \textbf{\bibinfo{volume}{49}},
  \bibinfo{pages}{1659} (\bibinfo{year}{1988}).

\bibitem[{\citenamefont{Xu et~al.}(2007)\citenamefont{Xu, Szulczewski, LeClair,
  Navarrete, Schad, Miao, Guo, and Gupta}}]{XuAPL2007}
\bibinfo{author}{\bibfnamefont{W.}~\bibnamefont{Xu}},
  \bibinfo{author}{\bibfnamefont{G.~J.} \bibnamefont{Szulczewski}},
  \bibinfo{author}{\bibfnamefont{P.}~\bibnamefont{LeClair}},
  \bibinfo{author}{\bibfnamefont{I.}~\bibnamefont{Navarrete}},
  \bibinfo{author}{\bibfnamefont{R.}~\bibnamefont{Schad}},
  \bibinfo{author}{\bibfnamefont{G.}~\bibnamefont{Miao}},
  \bibinfo{author}{\bibfnamefont{H.}~\bibnamefont{Guo}}, \bibnamefont{and}
  \bibinfo{author}{\bibfnamefont{A.}~\bibnamefont{Gupta}},
  \bibinfo{journal}{Appl. Phys. Lett.} \textbf{\bibinfo{volume}{90}},
  \bibinfo{pages}{072506} (\bibinfo{year}{2007}).

\bibitem[{\citenamefont{Ruden and Smith}(2004)}]{RudenSmithJAP2004}
\bibinfo{author}{\bibfnamefont{P.~P.} \bibnamefont{Ruden}} \bibnamefont{and}
  \bibinfo{author}{\bibfnamefont{D.~L.} \bibnamefont{Smith}},
  \bibinfo{journal}{J. Appl. Phys.} \textbf{\bibinfo{volume}{95}},
  \bibinfo{pages}{4898} (\bibinfo{year}{2004}).

\bibitem[{\citenamefont{Dediu et~al.}(2008)\citenamefont{Dediu, Hueso,
  Bergenti, Riminucci, Borgatti, Graziosi, Newby, Casoli, Jong, Taliani
  et~al.}}]{DediuPRB2008}
\bibinfo{author}{\bibfnamefont{V.}~\bibnamefont{Dediu}},
  \bibinfo{author}{\bibfnamefont{L.~E.} \bibnamefont{Hueso}},
  \bibinfo{author}{\bibfnamefont{I.}~\bibnamefont{Bergenti}},
  \bibinfo{author}{\bibfnamefont{A.}~\bibnamefont{Riminucci}},
  \bibinfo{author}{\bibfnamefont{F.}~\bibnamefont{Borgatti}},
  \bibinfo{author}{\bibfnamefont{P.}~\bibnamefont{Graziosi}},
  \bibinfo{author}{\bibfnamefont{C.}~\bibnamefont{Newby}},
  \bibinfo{author}{\bibfnamefont{F.}~\bibnamefont{Casoli}},
  \bibinfo{author}{\bibfnamefont{M.~P.~D.} \bibnamefont{Jong}},
  \bibinfo{author}{\bibfnamefont{C.}~\bibnamefont{Taliani}},
  \bibnamefont{et~al.}, \bibinfo{journal}{Phys. Rev. B}
  \textbf{\bibinfo{volume}{78}}, \bibinfo{eid}{115203} (\bibinfo{year}{2008}).

\bibitem[{\citenamefont{Hill et~al.}(1998)\citenamefont{Hill, Rajagopal, Kahn,
  and Hu}}]{HillAPL1998}
\bibinfo{author}{\bibfnamefont{I.~G.} \bibnamefont{Hill}},
  \bibinfo{author}{\bibfnamefont{A.}~\bibnamefont{Rajagopal}},
  \bibinfo{author}{\bibfnamefont{A.}~\bibnamefont{Kahn}}, \bibnamefont{and}
  \bibinfo{author}{\bibfnamefont{Y.}~\bibnamefont{Hu}}, \bibinfo{journal}{Appl.
  Phys. Lett.} \textbf{\bibinfo{volume}{73}}, \bibinfo{pages}{662}
  (\bibinfo{year}{1998}).

\bibitem[{\citenamefont{Kahn et~al.}(2003)\citenamefont{Kahn, Koch, and
  Gao}}]{KahnPolymer2003}
\bibinfo{author}{\bibfnamefont{A.}~\bibnamefont{Kahn}},
  \bibinfo{author}{\bibfnamefont{N.}~\bibnamefont{Koch}}, \bibnamefont{and}
  \bibinfo{author}{\bibfnamefont{W.}~\bibnamefont{Gao}}, \bibinfo{journal}{J.
  Polym. Sci. Part B: Polym. Phys.} \textbf{\bibinfo{volume}{41}},
  \bibinfo{pages}{2529} (\bibinfo{year}{2003}).

\bibitem[{\citenamefont{Tiba et~al.}(2006)\citenamefont{Tiba, de~Jonge,
  Koopmans, and Jonkman}}]{TibaJAP2006}
\bibinfo{author}{\bibfnamefont{M.~V.} \bibnamefont{Tiba}},
  \bibinfo{author}{\bibfnamefont{W.~J.~M.} \bibnamefont{de~Jonge}},
  \bibinfo{author}{\bibfnamefont{B.}~\bibnamefont{Koopmans}}, \bibnamefont{and}
  \bibinfo{author}{\bibfnamefont{H.~T.} \bibnamefont{Jonkman}},
  \bibinfo{journal}{J. Appl. Phys.} \textbf{\bibinfo{volume}{100}},
  \bibinfo{eid}{093707} (\bibinfo{year}{2006}).

\bibitem[{\citenamefont{Popinciuc et~al.}(2006)\citenamefont{Popinciuc,
  Jonkman, and van Wees}}]{PopinciucJAP2006}
\bibinfo{author}{\bibfnamefont{M.}~\bibnamefont{Popinciuc}},
  \bibinfo{author}{\bibfnamefont{H.~T.} \bibnamefont{Jonkman}},
  \bibnamefont{and} \bibinfo{author}{\bibfnamefont{B.~J.} \bibnamefont{van
  Wees}}, \bibinfo{journal}{J. Appl. Phys.} \textbf{\bibinfo{volume}{100}},
  \bibinfo{eid}{093714} (\bibinfo{year}{2006}).

\bibitem[{\citenamefont{Zhan et~al.}(2007)\citenamefont{Zhan, Bergenti, Hueso,
  Dediu, de~Jong, and Li}}]{ZhanPRB2007}
\bibinfo{author}{\bibfnamefont{Y.~Q.} \bibnamefont{Zhan}},
  \bibinfo{author}{\bibfnamefont{I.}~\bibnamefont{Bergenti}},
  \bibinfo{author}{\bibfnamefont{L.~E.} \bibnamefont{Hueso}},
  \bibinfo{author}{\bibfnamefont{V.}~\bibnamefont{Dediu}},
  \bibinfo{author}{\bibfnamefont{M.~P.} \bibnamefont{de~Jong}},
  \bibnamefont{and} \bibinfo{author}{\bibfnamefont{Z.~S.} \bibnamefont{Li}},
  \bibinfo{journal}{Phys. Rev. B} \textbf{\bibinfo{volume}{76}},
  \bibinfo{eid}{045406} (\bibinfo{year}{2007}).

\bibitem[{\citenamefont{Zhan et~al.}(2008)\citenamefont{Zhan, de~Jong, Li,
  Dediu, Fahlman, and Salaneck}}]{ZhanPRB2008}
\bibinfo{author}{\bibfnamefont{Y.~Q.} \bibnamefont{Zhan}},
  \bibinfo{author}{\bibfnamefont{M.~P.} \bibnamefont{de~Jong}},
  \bibinfo{author}{\bibfnamefont{F.~H.} \bibnamefont{Li}},
  \bibinfo{author}{\bibfnamefont{V.}~\bibnamefont{Dediu}},
  \bibinfo{author}{\bibfnamefont{M.}~\bibnamefont{Fahlman}}, \bibnamefont{and}
  \bibinfo{author}{\bibfnamefont{W.~R.} \bibnamefont{Salaneck}},
  \bibinfo{journal}{Phys. Rev. B} \textbf{\bibinfo{volume}{78}},
  \bibinfo{eid}{045208} (\bibinfo{year}{2008}).

\end{thebibliography}
\end{document}